**A mixed integer programming approach to minibeam aperture optimization for multi-collimator proton minibeam radiotherapy**

Nimita Shinde, Weijie Zhang, Yuting Lin, and Hao Gao[*]

Department of Radiation Oncology, University of Kansas Medical Center, USA

[*] Corresponding author:

Hao Gao, Department of Radiation Oncology, University of Kansas Medical Center, USA.

Email address: hgao2@kumc.edu

**Acknowledgments**

The authors are very thankful for the valuable comments from reviewers. This research is partially supported by the NIH grants No. R37CA250921, R01CA261964, and a KUCC physicist-scientist recruiting grant.

**Conflict of Interest Statement**

None.

**Ethical Statement:** This research was carried out under Human Subject Assurance Number 00003411 for University of Kansas in accordance with the principles embodied in the Declaration of Helsinki and in accordance with local statutory requirements. Consent was given for publication by the participants of this study.

**Abstract.**

**Background:** Multi-collimator proton minibeam radiation therapy (MC-pMBRT) has recently emerged as a versatile technique for dose shaping, enabling the formation of peak-valley dose patterns in organs-at-risk (OAR) while maintaining a uniform dose distribution in tumor targets. MC-pMBRT leverages a set of generic multi-slit collimators (MSC) with varying center-to-center distances. However, the current method for minibeam aperture optimization (MAO), i.e., the selection of MSC per beam angle to optimize plan quality, is manual and heuristic, resulting in computational inefficiencies and no guarantee of optimality.

**Purpose:** This work introduces a novel mixed integer programming (MIP) approach to MAO for optimizing MC-pMBRT plan quality.

**Methods:** The proposed MIP approach jointly optimizes dose distributions, peak-to-valley dose ratio (PVDR), and selects the optimal set of MSC per beam angle. The optimization problem includes decision variables for MSC selection per beam angle and spot weights. The proposed MIP approach is a two-step process: in the first step, the binary variables are optimally determined to select MSC for each beam angle; in the second step, the continuous variables are solved to determine the spot weights. Both steps utilize iterative convex relaxation and the alternating direction method of multipliers (ADMM) to solve the optimization problems efficiently.

**Results:** The proposed MIP method to solve MAO (MIP-MAO) was validated against the conventional heuristic method (CONV) for MC-pMBRT treatment planning. Results indicate that MIP-MAO enhances the conformity index (CI) for the target and improves PVDR for OAR. For instance, in a head-and-neck case, CI improved from 0.61 (CONV) to 0.70 (MIP-MAO); in an abdomen case, CI improved from 0.78 (CONV) to 0.83 (MIP-MAO). Additionally, MIP-MAO reduced mean doses in the body and OAR.

**Conclusions:** A novel MIP approach for MAO in MC-pMBRT is presented, showing demonstrated improvements in plan quality and PVDR compared to the heuristic method.

# 1. Introduction

Proton minibeam radiation therapy (pMBRT) [1, 2] is an emerging proton modality that delivers sub-millimeter proton beams (minibeams) spaced a few millimeters apart using multi-slit collimators (MSC) [1,3]. A key aspect of pMBRT is minibeam aperture optimization (MAO), i.e., the problem of selecting the optimal MSC with appropriate center-to-center (ctc) distances for each beam angle. This selection is crucial because the peak-to-valley dose ratio (PVDR) in pMBRT is highly sensitive to minibeam spacing [3, 4]. Optimizing MAO directly impacts dose conformity and the ability to achieve high PVDR in organs-at-risk (OAR) while maintaining a uniform dose distribution in the tumor.

A novel multi-collimator pMBRT method (MC-pMBRT) [5] utilizes MSC with varying ctc distances for each beam angle, and jointly optimizes both the PVDR for OAR and the dose distribution. This approach offers a significant advantage over conventional single-collimator methods by ensuring sufficient PVDR in normal tissues while preserving dose uniformity within the target. However, the current method for selecting MSC is heuristic, leading to a possibly suboptimal solution.

The challenge of optimal MSC selection arises from its combinatorial nature, resulting in an exponentially large solution space. For example, if three MSC options (ctc distances of 3mm, 4mm, and 5mm) are available across four beam angles (45º, 135º, 225º, 315º), there are 81 possible MSC combinations. This complexity makes mixed-integer programming (MIP) a promising approach, using binary decision variables to determine the optimal MSC configuration. MIP has been widely explored in radiotherapy [6-11] particularly for beam angle and beam orientation optimization [6, 8-10], supported by advancements in computational tools. Research on MIP approaches typically involves developing models and solving them using (i) off-the-shelf solvers [6], (ii) meta-heuristic methods [10] or (iii) problem-specific techniques such as column generation [10]. However, to the best of our knowledge, MIP has not yet been applied to MAO in pMBRT, presenting a novel opportunity for optimization in this domain.

To address the limitations of heuristic MSC selection and leveraging the advantages of MIP, this work proposes a MIP-based approach to MAO (MIP-MAO), enabling optimization of MSC selection for MC-pMBRT. By integrating MIP, the need for exhaustive enumeration of all possible MSC combinations is eliminated, reducing computational complexity while identifying near-optimal solutions efficiently. With a modest computational cost, MIP-MAO offers the potential for significantly improved dose planning, maximizing PVDR and enhancing treatment outcomes.

## 2. Methods

### 2.1. Joint dose and PVDR optimization (JDPO) in MC-pMBRT

MC-pMBRT [5] jointly optimizes dose and PVDR by solving the JDPO optimization problem

$$\min_{x_j} f(d) + \sum_{k=1}^{K} w^k g(d^k)$$
$$\text{s.t. } x_j \in \{0\} \cup [G, +\infty) \; \forall j = 1, \dots, B,$$
$$d = \sum_{j=1}^{B} A_j x_j, \quad (1)$$
$$d^k = I_k d, k = 1, \dots K.$$

In Eq. (1), $x_j$ represents the spot intensity vector for each beam angle $j = 1, \dots, B$ to be optimized, $A_j$ represents the dose influence matrix for each beam angle $j$, and $G$ is the minimum monitor unit (MMU) threshold value. The first constraint in Eq. (1) is an MMU constraint [12, 13] for $x_t$ that ensures plan deliverability. The second constraint defines the dose distribution. The third constraint defines the dose distributions $d^k$ at each $k = 1, \dots, K$ beam-eye-view 2D dose plane located at specific depth from the beam angle. At each beam angle, at least one beam-eye-view is considered for PVDR optimization.

The first term in the objective, $f(d)$, defines the dose distribution to be optimized and is given as

$$f(d) = \sum_{i=1}^{N_1} \frac{w_1}{n_i} ||d_{\Omega_{1i}} - b_{1i}||_2^2 + \sum_{i=1}^{N_2} \frac{w_2}{n_i} ||d_{\Omega_{2i}} - b_{2i}||_2^2 + \frac{w_3}{n} ||d_{\Omega_3} - b_3||_2^2.$$

We now briefly describe the three terms in the objective function $f(d)$.

- The first term defines $N_1$ least square error terms for the difference between the actual dose distribution $d_{\Omega_{1i}}$ and the desirable dose $b_{1i}$ for the target as well as OAR. Here, $\Omega_{1i}$ is the set of active indices for target/OAR at which the least square term is defined.

- The second term describes $N_2$ dose volume histogram (DVH)-max constraints [14, 15] for OAR. The DVH-max constraint defined for any OAR $i$ states that at most $p$ fraction of the total voxels in the OAR should receive a dose larger than $b_{2i}$. A commonly used technique to define the DVH-max constraint involves defining the set $\Omega_{2i}$ of indices that violate the constraint. Let $d'$ be the dose distribution $d$ sorted in descending order and let $n_i$ be the number of voxels in OAR $i$. Then, $\Omega_{2i} = \{j|j \geq p \times n_i\}$ if $d'_{p \times n_i} \geq b_{2i}$, i.e., if the DVH-max constraint is violated. The second term in $f(d)$ then defines the least square error between the dose $d_{\Omega_{2i}}$ (based on the active index set), and the DVH-max dose value $b_{2i}$.

- The third term in $f(d)$ describes DVH-min constraint [14, 15] for the target. The DVH-min constraint states that at least $p$ fraction of the total voxels in the target should receive a dose larger than $b_3$. Let $d'$ be the dose distribution $d$ sorted in descending order and let $n$ be the number of voxels in the target. Then, $\Omega_3 = \{j|j \leq p \times n\}$ if $d'_{p \times n} \leq b_3$, i.e., the DVH-min constraint is violated. Thus, the third term in $f(d)$ defines the least square error between the dose $d_{\Omega_3}$ for the indices that violate the DVH-min constraint and minimum dose value $b_3$.

The term, $g(d^k)$, in the objective of the Eq. (1) is the PVDR regularization term and is defined as $g(d^k) = ||d^k||_1 - w_T||Td^k||_1$ [5], where $T$ is the linear operator that defines the point-wise difference between dose received by adjacent voxels in 2D dose plane $d^k$.

The solution to Eq. (1) is a spot intensity vector that optimizes both dose distribution and PVDR. The dose influence matrix $A_j$ in the JDPO optimization model corresponds to the MSC chosen a priori for the beam angle $j$. Section 2.2 provides an optimization model that uses MIP to instead choose MSC optimally for each beam angle.

*2.2 MIP-MAO method for minibeam aperture optimization*

In contrast to JDPO optimization model [5] Eq. (1) (where the choice of MSC for each beam angle is fixed before the model is defined), MIP-MAO aims to optimally select MSC for every beam angle. Assume that $C$ different MSC of varying ctc distances are available to be used for all $B$ beam angles, and define a binary decision variable $y_{ij}$ $\forall\, i = 1, \ldots, C, j = 1, \ldots, B$ such that $y_{ij} = 1$ if MSC $i$ is selected for use at beam angle $j$. Then, MIP formulation for MAO is

$$\min_{x_j, y} f(d) + \sum_{k=1}^{K} w_k g(d^k)$$
$$\text{s.t. } x_j \in \{0\} \cup [G, +\infty) \quad \forall\, j = 1, \ldots, B,$$
$$d = \sum_{j=1}^{B} \left( \sum_{i=1}^{C} A_{ij} y_{ij} \right) x_j, \tag{2}$$
$$d_k = I_k d, \quad \forall\, k = 1, \ldots K$$
$$\sum_{i=1}^{C} y_{ij} = 1 \quad \forall\, j = 1, \ldots, B,$$
$$y_{ij} = \{0,1\}, \quad \forall\, i = 1, \ldots, C, j = 1, \ldots, B.$$

The two terms in the objective of Eq. (2) are described in Section 2.1, and are similar to the objective function in Eq. (1). Furthermore, the first and third constraint in Eq. (2) are the same as in Eq. (1). The second constraint defines the dose distribution based on (i) the MSC $i$ chosen by $y_{ij}$ and (ii) spot intensity vector $x_j$ at each beam angle $j$. Finally, the fourth constraint in the model ensures that exactly one MSC is chosen for every beam angle.

The proposed MIP-MAO method follows a two-step process. In the first step, Eq. (2) is solved to determine an optimal combination of MSC for each beam angle. The decision variables $x_j$'s in the output of Eq. (2) provide spot intensities that optimize dose plan. However, to reduce computational complexity, Eq. (2) is solved approximately, yielding a near-optimal MSC selection. In the second step, there remains an opportunity to further refine the spot intensities. Thus, Eq. (2) is solved again using the MSC configuration obtained from the first step to generate the final set of optimized spot intensities.

*2.3 Solution algorithm for MIP-MAO method*

*First step of MIP-MAO:* Solving Eq. (2) in the first step of MIP-MAO method begins with introducing auxiliary variables and relaxing the binary constraint on the $y_{ij}$ variables. Then, Eq. (2) is re-written as

$$\min_{x_j, y} f\left(\sum_{j=1}^{B}\left(\sum_{i=1}^{C} A_{ij} y_{ij}\right) x_j\right) + \sum_{k=1}^{K}(w_k ||u_k||_1 - w_k w_T ||v_k||_1)$$

$$s.t. \quad z_j \in \{0\} \cup [G, +\infty) \quad \forall j = 1, \ldots, B,$$

$$z_j = x_j, \quad \forall j = 1, \ldots, B,$$

$$u_k = I_k \sum_{j=1}^{B}\left(\sum_{i=1}^{C} A_{ij} y_{ij}\right) x_j, \quad \forall k = 1, \ldots K \quad (3)$$

$$v_k = TI_k \sum_{j=1}^{B}\left(\sum_{i=1}^{C} A_{ij} y_{ij}\right) x_j, \quad \forall k = 1, \ldots K$$

$$\sum_{i=1}^{C} y_{ij} = 1 \quad \forall j = 1, \ldots, B.$$

Eq. (3) can now be solved using iterative convex relaxation (ICR) method [16, 17] and alternating direction method of multipliers (ADMM) method [18, 19], which has been successfully used to solve inverse optimization problems [20-29]. The iterative method involves updating the active index sets for the DVH constraints followed by updating each decision variable in the problem sequentially while keeping other variables fixed. To use ADMM method, define the augmented Lagrangian for Eq. (3) as

$$\min_{x_j, y} f\left(\sum_{j=1}^{B}\left(\sum_{i=1}^{C} A_{ij} y_{ij}\right) x_j\right) + \sum_{k=1}^{K}(w_k ||u_k||_1 - w_k w_T ||v_k||_1) + \frac{\mu_1}{2}\sum_{j=1}^{B} ||z_j - x_j + \lambda_{1j}||_2^2$$

$$+ \frac{\mu_2}{2}\sum_{j=1}^{B}||\sum_{i=1}^{C} y_{ij} - 1 + \lambda_{2j}||_2^2 + \frac{\mu_3}{2}\sum_{k=1}^{K} ||u_k - I_k \sum_{j=1}^{B}\left(\sum_{i=1}^{C} A_{ij} y_{ij}\right) x_j + \lambda_{3k}||_2^2$$

$$+ \frac{\mu_4}{2}\sum_{k=1}^{K} ||v_k - TI_k \sum_{j=1}^{B}\left(\sum_{i=1}^{C} A_{ij} y_{ij}\right) x_j + \lambda_{4k}||_2^2 \quad (4)$$

$$s.t. \quad z_j \in \{0\} \cup [G, +\infty) \quad \forall j = 1, \ldots, B.$$

Algorithm 1 provides a brief outline of the optimization method for solving Eq. (4). The detailed explanation of the steps involved in updating the decision variables (Step 4b of Algorithm 1) are given in Appendix A. The output of Algorithm 1, i.e., the solution to Eq. (4) is the spot intensity vector $x_j$ and fractional variable $y_{ij}$. Since a binary value of $y$ is needed to determine the MSC to use at every beam angle, the output $y_{ij}$ of Algorithm 1 can be projected onto the constraint set $\{y \in \{0,1\}, \sum_{i=1}^{C} y_{ij} = 1\}$. The resulting binary solution provides the information about the MSC to select at every beam angle.

*Second step of MIP-MAO:* Once the nearly optimal choice of MSC is obtained from the first step of the method, Eq. (2) is solved again, this time with $y_{ij}$ fixed. The solution methodology follows a similar approach to that described in Algorithm 1 for the first step of the MIP-MAO method, with the key distinction that $y_{ij}$ remains fixed.

---

**Algorithm 1: Optimization method for solving Eq. (4)**

1. **Input:** Choose parameters $\mu_1, \ldots, \mu_4, w_1, w_2, w_3, w_k, w_T$
2. Initialization: Randomly initialize $x_j, y$. Choose number of iterations $T$.
3. Set $z_j = x_j$, $u_k - I_k \sum_{j=1}^{B}\left(\sum_{i=1}^{C} A_{ij} y_{ij}\right) x_j$, $v_k - TI_k \sum_{j=1}^{B}\left(\sum_{i=1}^{C} A_{ij} y_{ij}\right) x_j$, $\lambda_{1j} = \lambda_{2j} = \lambda_{3k} = \lambda_{4k} = 0$.
4. For $t = 1, \ldots, T$
   a. Find active index sets $\Omega_{1i}, \Omega_{2i}, \Omega_3$ for DVH constraints as described in Section 2.1.
   b. Update primal variables $x_j, z_j, y_{ij}, u_k, v_k$ by fixing all variables except one and solving the resulting minimization problem.
   c. Update dual variables as follows:

$$\lambda_{1j} = \lambda_{1j} + z_j - x_j$$

$$\lambda_{2j} = \lambda_{2j} + \sum_{i=1}^{C} y_{ij} - 1$$

$$\lambda_{3k} = \lambda_{3k} + u_k - I_k \sum_{j=1}^{B}\left(\sum_{i=1}^{C} A_{ij} y_{ij}\right) x_j$$

$$\lambda_{4k} = \lambda_{4k} + v_k - TI_k \sum_{j=1}^{B}\left(\sum_{i=1}^{C} A_{ij} y_{ij}\right) x_j.$$

5. **Output:** $x_j, y$

---

*2.4 Materials*

The advantages of the proposed MIP-MAO method over the heuristic method (CONV) for MC-pMBRT treatment planning are demonstrated through three clinical test cases: abdomen, lung, and head-and-neck (HN). The beam angles used for the abdomen and lung cases are (0º, 120º, 240º), while for the HN case, they are (45º, 135º, 225º, 315º). For the abdomen and lung cases, three MSC options with center-to-center (ctc) distances of 3 mm, 5 mm, and 7 mm are considered for each beam angle. In the HN case, the available MSC choices have ctc distances of 3 mm, 4 mm, and 5 mm. The slit width for all MSC is 0.4 mm. Planning target volume (PTV) dose plans are evaluated using both CONV (reference model) and MIP-MAO (proposed model). The dose influence matrix for each beam angle and MSC configuration is generated using MatRad [30], assuming a spot width of 0.4 mm on a 1×1×3 mm³ dose grid.

To ensure comparability, all plans are normalized so that at least 95% of the PTV receives 100% of the prescription dose. Dose plan quality is assessed using the Conformity Index (CI) and the maximum dose delivered to the tumor ($D_{max}$). The CI is defined as CI=$V_{100}^2/(V \times V'_{100})$, $V_{100}$ is the PTV that receives at least 100% of the prescription dose, $V$ is the PTV volume, and $V'_{100}$ is the total volume that receives at least 100% of the prescription dose. The normalized $D_{max}$ is calculated as as $(D/D_p) \times 100\%$, where D is the maximum dose delivered to the tumor, and $D_p$ is the prescription dose. Furthermore, to determine the quality of PVDR optimization, PVDR is calculated as $D_{10}/D_{80}$, where $D_{10}$ and $D_{80}$ are the doses delivered to at least 10% and 80% of the entire volume respectively [5, 31].

## 3. Results

*3.1 Optimal choice of MSC for each clinical test case*

For the HN case, the CONV method [5] for MC-pMBRT uses MSC with ctc distances 3 mm, 5 mm, 5 mm, 3 mm at beam angles (45º, 135º, 225º, 315º) respectively. The proposed method, MIP-MAO, chooses MSC with ctc distances 3 mm, 4 mm, 4 mm, 3 mm for the same beam angles. For the abdomen case, the CONV method uses MSC with ctc distances of 3 mm, 3 mm, and 7 mm at beam angles (0º, 120º, 240º) respectively. The MIP-MAO method, however, selects MSC with ctc distances of 3 mm across all beam angles. Finally, for the lung case, the CONV method uses MSC with ctc distances of 3 mm, 5 mm, and 7 mm at beam angles (0º, 120º, 240º) respectively while MIP-MAO selects MSC with ctc distances 3 mm, 5mm, 5 mm for the three beam angles (0º, 120º, 240º).

*3.2 Comparison of dose plan quality*

Tables 1-3 indicate an increase in the conformity index (CI) when using the MIP-MAO method across all cases. Notably, the HN case exhibits a significant improvement in CI from 0.615 (CONV) to 0.703 (MIP-

MAO), while the abdomen case improves from 0.782 (CONV) to 0.837 (MIP-MAO). Additionally, in all cases, the mean dose delivered to OAR and the maximum dose delivered to the target remain comparable between the two methods.

*3.3 Comparison of PVDR*

The performance of the CONV and MIP-MAO methods is further assessed in terms of PVDR optimization. As shown in Tables 1-3, PVDR values across nearly all 2D dose planes are higher for the MIP-MAO method. For the HN case (Table 1), PVDR is calculated at four 2D dose planes, with an increase observed in three of them. A particularly notable improvement is seen at the 135º dose plane, where PVDR rises from 7.686 (CONV) to 10.66 (MIP-MAO). In the abdomen case, all three 2D dose planes exhibit increased PVDR values with MIP-MAO. A substantial improvement is noted at the 240º beam angle, where PVDR increases from 4.979 (CONV) to 7.328 (MIP-MAO). For the lung case, the MIP-MAO method results in only a slight improvement in PVDR values compared to the CONV method.

Table 1: Comparison of CONV and MIP-MAO method for HN case. The value (3553) indicates that MSC with distances 3 mm, 5 mm, 5 mm, 3 mm respectively were used for beam angles (45º, 135º, 225º, 315º) respectively. PVDR at beam angles (45º, 135º, 225º, 315º) are calculated for 2D dose planes at the depth of (2.5 cm, 5 cm, 5 cm, 2.5 cm) respectively. Improved CI and PVDR values are highlighted using bold text.

| Quantity | CONV (3553) | MIP-MAO (3443) |
|---|---|---|
| Obj fn val | 18.394 | 18.119 |
| CI | 0.615 | **0.703** |
| $D_{max}$ | 124.55% | 125.21% |
| $D_{mean}$ (body) | 0.760% | 0.738% |
| $D_{mean}$ (larynx) | 5.501% | 5.445% |
| $D_{mean}$ (mandible) | 6.204% | 6.201% |
| $D_{mean}$ (oral) | 4.645% | 4.710% |
| $D_{mean}$ (45º) | 1.143 | 1.139 |
| PVDR (45º) | 11.687 | **11.692** |
| $D_{mean}$ (135º) | 0.881 | 0.854 |
| PVDR (135º) | 7.686 | **10.66** |
| $D_{mean}$ (225º) | 0.865 | 0.781 |
| PVDR (225º) | 9.267 | **9.749** |
| $D_{mean}$ (315º) | 1.084 | 1.081 |
| PVDR (315º) | 13.436 | 13.357 |

Table 2: Comparison of CONV and MIP-MAO method for abdomen case. The value (337) indicates that MSC with distances 3 mm, 3 mm, 7 mm respectively were used for beam angles (0º, 120º, 240º) respectively. PVDR at beam angles (0º, 120º, 240º) are calculated for 2D dose planes at the depth of (4 cm, 2 cm, 9 cm) respectively. Improved CI and PVDR values are highlighted using bold text.

| Quantity | CONV (337) | MIP-MAO (333) |
| --- | --- | --- |
| Obj fn val | 281.60 | 281.58 |
| CI | 0.782 | **0.837** |
| $D_{max}$ | 122.98% | 122.89% |
| $D_{mean}$ (body) | 4.120% | 4.114% |
| $D_{mean}$ (large bowel) | 18.281% | 18.305% |
| $D_{mean}$ (spinal cord) | 6.003% | 5.904% |
| $D_{mean}$ (left kidney) | 21.796% | 21.808% |
| $D_{mean}$ (0º) | 18.010% | 18.137 |
| PVDR (0º) | 15.374 | **15.41** |
| $D_{mean}$ (120º) | 15.075 | 15.154 |
| PVDR (120º) | 9.751 | **9.907** |
| $D_{mean}$ (240º) | 4.507 | 4.397 |
| PVDR (240º) | 4.979 | **7.328** |

Table 3: Comparison of CONV and MIP-MAO method for lung case. The value (357) indicates that MSC with distances 3 mm, 5 mm, 7 mm respectively were used for beam angles (0º, 120º, 240º) respectively. PVDR at beam angles (0º, 120º, 240º) are calculated for 2D dose planes at the depth of (3 cm, 7 cm, 12 cm) respectively. Improved CI and PVDR values are highlighted using bold text.

| Quantity | CONV (357) | MIP-MAO (355) |
| --- | --- | --- |
| Obj fn val | 10.69 | 10.631 |
| CI | 0.612 | **0.626** |
| $D_{max}$ | 135.53% | 136.62% |
| $D_{mean}$ (body) | 3.902% | 3.832% |
| $D_{mean}$ (lung) | 7.611% | 7.449% |
| $D_{mean}$ (heart) | 2.138% | 2.101% |
| $D_{mean}$ (eso) | 6.475% | 5.821% |
| $D_{mean}$ (0º) | 5.034 | 5.111 |
| PVDR (0º) | 13.969 | **13.993** |
| $D_{mean}$ (120º) | 3.908 | 3.995 |
| PVDR (120º) | 7.915 | **7.973** |
| $D_{mean}$ (240º) | 2.545 | 2.287 |
| PVDR (240º) | 7.949 | 6.138 |

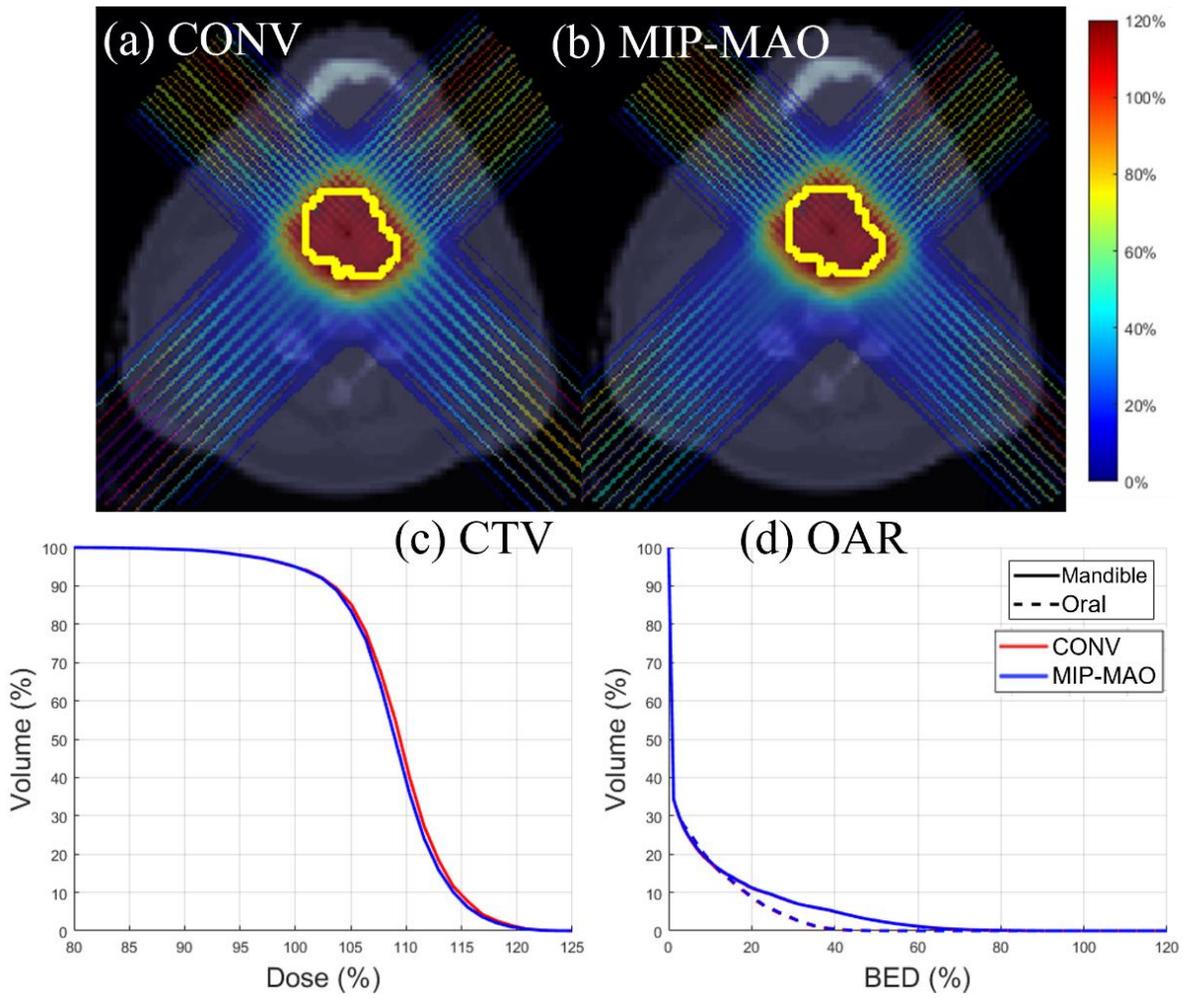

Figure 1. **HN**. (a), (b) Dose plots for CONV and MIP-MAO methods respectively, (c) DVH plot for the target, (d) DVH plot for OAR

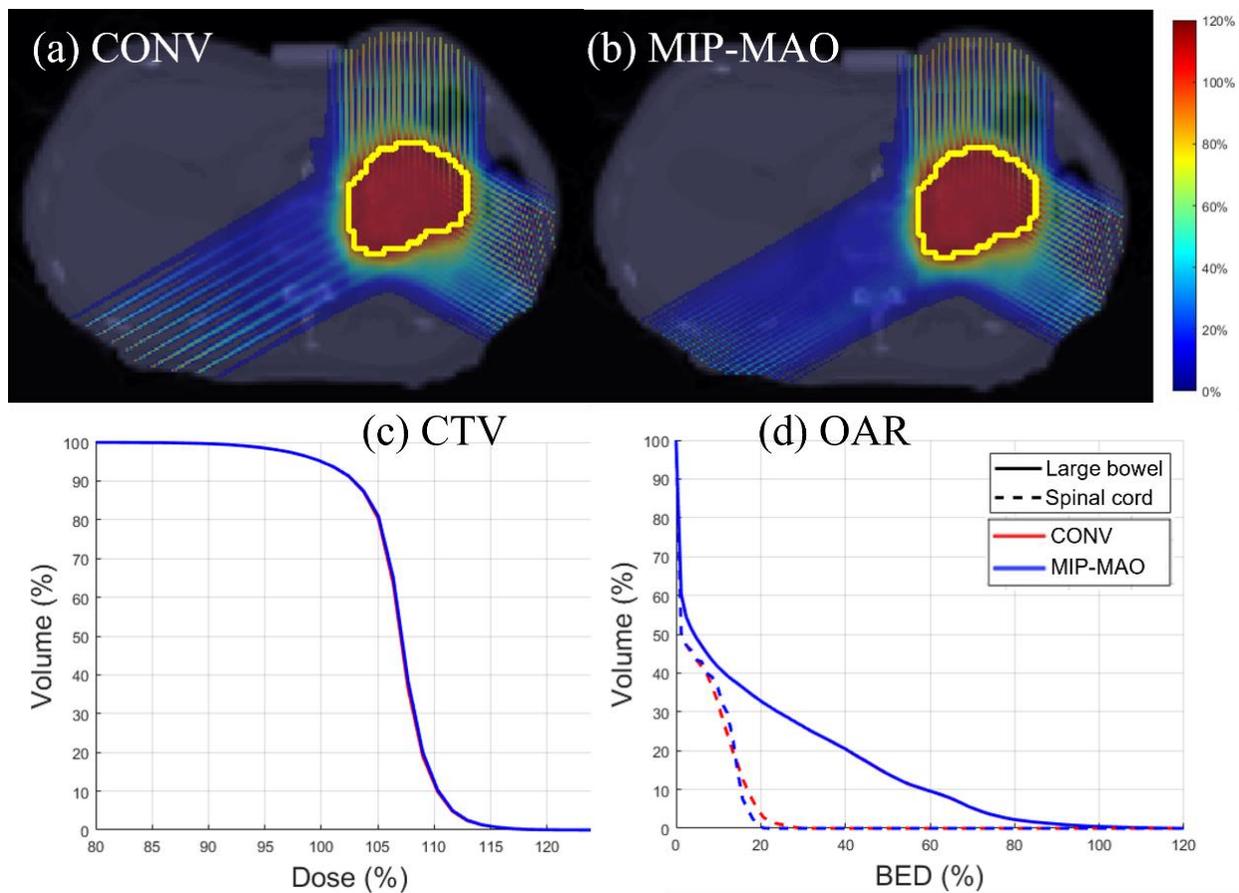

Figure 2. **Abdomen**. (a), (b) Dose plots for CONV and MIP-MAO methods respectively, (c) DVH plot for the target, (d) DVH plot for OAR

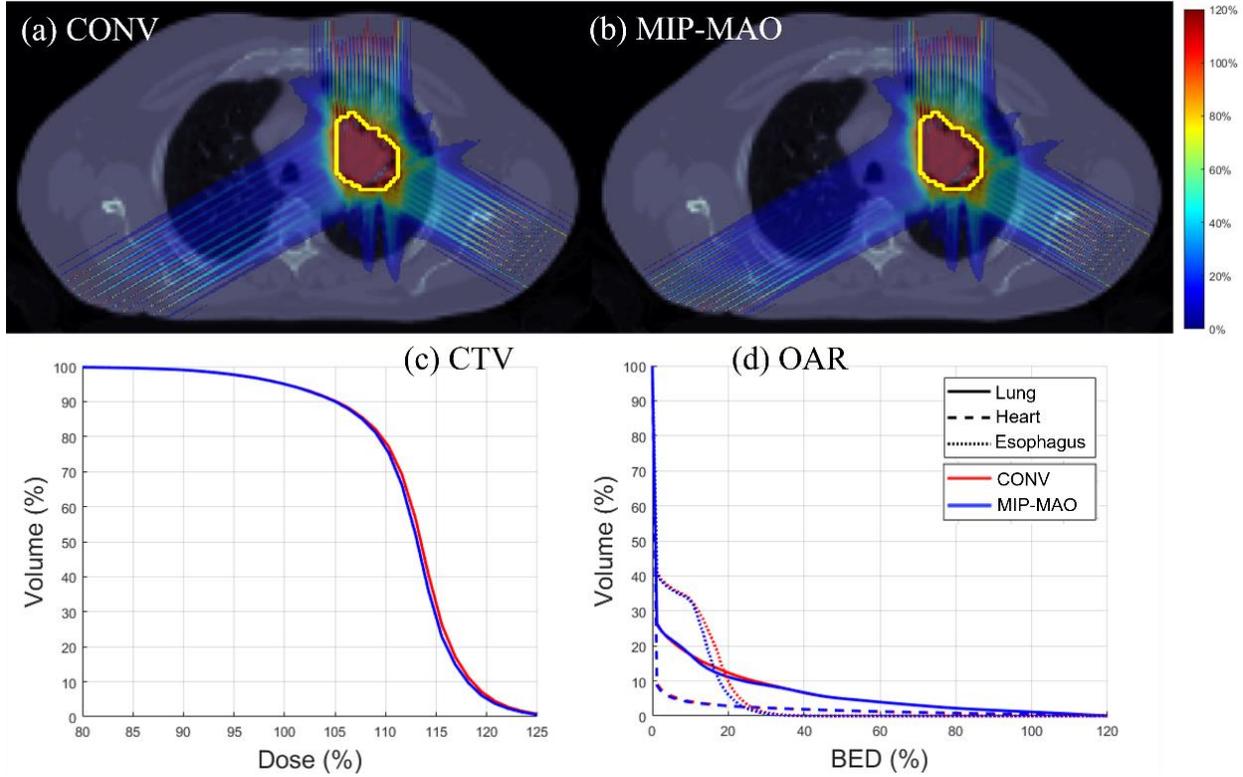

Figure 3. **Lung**. (a), (b) Dose plots for CONV and MIP-MAO methods respectively, (c) DVH plot for the target, (d) DVH plot for OAR

## 4. Discussion

This work presents an MIP approach for MAO in MC-pMBRT treatment planning. MIP provides a robust framework for solving complex optimization problems involving binary decisions that lead to exponentially large solution spaces. For example, in proton radiotherapy, energy layer optimization (ELO) seeks to minimize the number of energy layers used in dose delivery while maintaining treatment quality. This problem involves binary decisions, where the objective is to select a subset of energy layers from a large set of available layers. A MIP-based model can effectively optimize this selection process. The iterative optimization method PER [32] previously applied MIP to ELO, demonstrating improvements over conventional approaches that utilize all energy layers. However, further refinements in solution strategies could enhance computational efficiency and treatment outcomes.

Beyond MAO and ELO, other optimization challenges in radiotherapy, such as beam angle optimization [33] and LATTICE peak optimization [34], could also benefit from MIP modeling and efficient solution strategies. The results presented in this work suggest that even approximate solutions from an MIP model can outperform meta-heuristic approaches for MAO, highlighting the potential of MIP in radiotherapy optimization.

A limitation of the proposed MIP-MAO model is the increased computational burden. This arises primarily from (1) the additional step required to determine the optimal combination of MSCs, which introduces more variables into the optimization problem, and (2) the increased problem size due to the need for a dose influence matrix and additional decision variables for each MSC option. Despite this limitation, the use of MIP is justified by the observed improvements in dose plan quality and PVDR.

Overall, MIP provides a distinct advantage in solving MAO and holds promise for broader applications in radiation therapy.

## 5. Conclusion

In this work, a mixed integer programming approach is proposed to choose an optimal combination of MSC to use at every beam angle in pMBRT treatment planning. The optimization model was inspired by the JDPO model [5]. However, their model chose the MSC to use *a priori* before solving the optimization problem. Our proposed model was able to provide an optimal selection of MSC resulting in improved dose plan quality and PVDR.

## Appendix A: ICR and ADMM method (Algorithm 1) for solving the Augmented Lagrangian formulation Eq. (4)

In this section, a detailed explanation of Step 4b of Algorithm 1 (where the primal variables in Eq. (4) are updated) is provided.

1. Updating $x_j$: For each $j = 1, \ldots, B$, fix all variables except $x_j$ in Eq. (4). The resulting minimization problem is unconstrained in $x_j$. Thus, to solve the problem, take the first-order derivative of the objective function with respect to $x_j$ and set it to 0. The value of $x_j$ is obtained by finding the solution to the resulting linear system of equations.

2. Updating $z_j$: For each $j = 1, \ldots, B$, fix all variables except $z_j$ in Eq. (4). The resulting minimization problem has a closed form solution that can be defined by soft thresholding as follows:

$$z_j = \begin{cases} max\,(G, x_j - \lambda_{1j}), & if\ x_j - \lambda_{1j} \geq G/2 \\ 0, & otherwise. \end{cases}$$

3. Updating $y$: In Eq. (4), fix all variables except $y$. The resulting minimization problem is unconstrained in $y$. Similar to updating $x_j$, to find the value of $y$, take the first-order derivative of the objective function with respective to $y$, set it to 0, and solve the resulting linear system of equations.

4. Updating $u_k$: For each $k = 1, \ldots, K$, fix all variables except $u_k$. The resulting unconstrained minimization problem is defined as

$$\min_{u_k}\ w_k\,||u_k||_1 + \frac{\mu_3}{2}\,||u_k - I_k \sum_{j=1}^{B}\left(\sum_{i=1}^{C} A_{ij}\,y_{ij}\right)x_j + \lambda_{3k}||_2^2.$$

Then, the closed form solution to unconstrained optimization is defined as

$$u_k = sign\left(I_k \sum_{j=1}^{B}\left(\sum_{i=1}^{C} A_{ij}\,y_{ij}\right)x_j - \lambda_{3k}\right)$$

$$* max\left(abs\left(I_k \sum_{j=1}^{B}\left(\sum_{i=1}^{C} A_{ij}\,y_{ij}\right)x_j - \lambda_{3k}\right) - \frac{w_k}{\mu_3},\,0\right).$$

5. Updating $v_k$: The procedure to update $v_k$ is similar to updating $u_k$. For each $k = 1, \ldots, K$, fix all variables except $v_k$. The resulting minimization problem is

$$\min_{v_k}\ -w_T w_k\,||v_k||_1 + \frac{\mu_4}{2} \sum_{k=1}^{K}||v_k - TI_k \sum_{j=1}^{B}\left(\sum_{i=1}^{C} A_{ij}\,y_{ij}\right)x_j + \lambda_{4k}||_2^2.$$

The solution to $v_k$ is given as

$$v_k = sign\left(TI_k \sum_{j=1}^{B}\left(\sum_{i=1}^{C} A_{ij}\, y_{ij}\right) x_j - \lambda_{4k}\right)$$

$$* \, max\left(abs\left(TI_k \sum_{j=1}^{B}\left(\sum_{i=1}^{C} A_{ij}\, y_{ij}\right) x_j - \lambda_{4k}\right) + \frac{w_T w_k}{\mu_4}\right), 0).$$